\documentclass[aps,prl,superscriptaddress,twocolumn,nofootinbib,preprintnumbers]{revtex4-2}
\usepackage{amsmath}
\usepackage{graphicx}
\usepackage{subfigure}
\usepackage{amssymb}
\usepackage{xcolor}
\usepackage{multirow}
\usepackage{cancel}
\usepackage{color}
\usepackage{ulem}
\usepackage{listings}
\usepackage{xcolor}
\usepackage{float}
\usepackage{slashed}
\usepackage{wrapfig}
\usepackage{mathtools}
\usepackage[colorlinks,citecolor=blue]{hyperref}
\usepackage[T1]{fontenc}  
\usepackage[utf8]{inputenc}  
\usepackage{times}
\begin{document}
	
\title{Self-Interaction Bounds on Ultralight Dark Matter Couplings to Matter}
    \author{Mohammad Aghaie}
	\email{aghaie@het.phys.sci.osaka-u.ac.jp}
	\affiliation{Department of Physics, The University of Osaka, Toyonaka, Osaka 560-0043, Japan}
	
	\author{Shao-Ping Li}
	\email{Shaoping.Li@oeaw.ac.at}
    	\affiliation{Marietta Blau Institute for Particle Physics, Austrian Academy of Sciences, Dominikanerbastei 16, A-1010 Vienna, Austria}

\begin{abstract}
Ultralight dark matter (ULDM) couplings to matter fields and ULDM self-interactions are typically treated as independent probes. However, since the ULDM-matter couplings unavoidably induce self-interactions through quantum loop corrections, bounds on self-interacting ULDM from astrophysical and cosmological observations will also limit the coupling strength to matter. Applying this argument,  we find that self-interaction bounds can impose strong constraints on the linear ULDM couplings to neutrinos, excluding a large portion of parameter space that is widely considered for probing ULDM via  neutrino oscillation experiments. In addition,  the self-interaction bounds also limit the quadratic ULDM couplings to electrons and light quarks, which can become stronger than from the stringent test of  equivalence-principle violation. Our results  demonstrate that the extreme observational sensitivity of cosmic microwave background and structure formations to repulsive self-interactions can robustly translate into powerful constraints on the ULDM interactions with fundamental particles.
\end{abstract}

\maketitle
\preprint{OU-HET-1311} 
\textbf{\textit{Introduction}}. 
Ultralight dark matter (ULDM) has received increasing attention over the past years~\cite{Arias:2012az,Hui:2016ltb,Ferreira:2020fam,Antypas:2022asj}. Being made up of bosonic particles with masses far below the eV scale, ULDM behaves as cold DM on megaparsec scales, and can further  explain kiloparsec-scale problems via  quantum interference beyond the cold DM paradigm~\cite{Hu:2000ke,Schive:2014dra,Bar:2018acw,May:2021wwp}. 

In addition to its gravitational effects, experimental searches for ULDM rely on its interactions with stable Standard Model (SM) fields, such as photons, charged leptons, light hadrons, and neutrinos. Over the past decades, searches—particularly those based on violations of the equivalence principle (EP), including fifth-force tests and measurements of oscillating constants~\cite{Smith:1999cr,Schlamminger:2007ht,Damour:2010rp,Wagner:2012ui,Stadnik:2015kia,Touboul:2017grn}, as well as high-precision atomic clock comparisons sensitive to temporal variations of fundamental constants~\cite{VanTilburg:2015oza, Hees:2016gop, BACON:2020ubh, Sherrill:2023zah}—have constrained linear ULDM couplings to extremely weak levels. Nevertheless, new detection channels continue to be explored for probing regions beyond these limits. In particular, ULDM interactions with neutrinos motivate a variety of detection strategies based on neutrino oscillations~\cite{Berlin:2016woy,Krnjaic:2017zlz,Dev:2020kgz,Losada:2021bxx,Gherghetta:2023myo}. Due to the weakly interacting nature of neutrinos, this channel remains less constrained. Meanwhile, nonlinear (quadratic) ULDM couplings to SM particles have attracted increasing attention, as they can evade stringent fifth-force bounds and exhibit qualitatively distinct phenomenology~\cite{Hees:2018fpg,Banerjee:2022sqg,Bouley:2022eer}.

In addition to the probes based on ULDM-matter interactions, ULDM self-interactions provide a highly sensitive handle on its properties. They affect both cosmological and astrophysical observables. On galactic scales, ULDM forms solitonic cores whose size and mass scale inversely with ULDM mass and depend sensitively on the self-interaction. Repulsive interactions enlarge these cores and can induce vortex formation~\cite{Rindler-Daller:2011afd,Dmitriev:2021utv}, whereas attractive interactions shrink them and may trigger instabilities~\cite{Arakawa:2024lqr,Chavanis:2011zm,Chavanis:2016dab}. Additional constraints arise from tidal disruption of solitons~\cite{Glennon:2022huu}, dynamical friction relevant for galaxy and supermassive black hole formation~\cite{Glennon:2023gfm}, and modifications to halo density profiles in galactic centers~\cite{Chakrabarti:2022owq}. Altogether, these effects impose extremely stringent bounds on the effective quartic coupling, reaching values about $\mathcal{O}(10^{-96})$ for ULDM mass around $10^{-22}$ eV.

Crucially, interactions between ULDM and SM fields invoked for experimental probes unavoidably generate effective self-interactions through quantum loop corrections, which are subject to the stringent bounds discussed above. In this Letter, we identify a simple mapping that translates these bounds into constraints on ULDM–matter couplings. In the absence of fine-tuned cancellations between the bare and loop-induced contributions, this mapping yields leading constraints across a wide range of ULDM mass. We find that, for linear couplings to neutrinos, 
the bounds derived from self-interactions based on cosmic microwave background (CMB) measurements dominate over existing limits. With the future CMB experiments, the linear ULDM-neutrino coupling can be further limited, exceeding  the projected sensitivity of upcoming neutrino oscillation experiments such as JUNO~\cite{JUNO:2015zny} and DUNE~\cite{DUNE:2015lol}. For nonlinear couplings, the impact is even more pronounced: the mapping can yield leading constraints for couplings to electrons and light quarks, which are stronger than those based on EP violation experiments such as E\"ot-Wash~\cite{Schlamminger:2007ht,Wagner:2012ui} and MICROSCOPE~\cite{Touboul:2017grn}. Our results apply broadly to both linear and nonlinear couplings of ULDM to fermions and bosons, providing a powerful and underexploited probe of ULDM interactions.

\textbf{\textit{Loop-induced self-coupling}}.
We consider a real scalar ULDM field $\phi$ interacting with SM fermions. While we focus on linear and quadratic couplings as representative examples, the underlying argument is more general. All interactions that contribute to the scalar effective potential generically induce loop-level self-interactions, allowing bounds on ULDM self-interactions to constrain the corresponding microscopic couplings. The electron and neutrino cases therefore serve as illustrative benchmarks of a broader class of ULDM interactions.

For linear interactions we use the Yukawa-like parametrization
\begin{align}
    \mathcal{L}_{\rm lin} \supset - y_f \,\phi\,\bar f f ,
    \label{lag:linear}
\end{align}
where $y_f$ is the dimensionless coupling.  For quadratic interactions we follow the standard convention used in equivalence-principle and cosmological analyses~\cite{Damour:2010rp,Stadnik:2015kia}, in which the scalar background induces quadratic variations of the fermion masses.  The relevant low-energy interactions are
\begin{align}
    \mathcal{L}_{\rm quad}
    \supset
    -2\pi \frac{\phi^2}{M_{\rm Pl}^2} \left(d_{m_f}^{(2)}+\gamma_m d_g^{(2)}\right)
        m_f \bar f f\,,
    \label{lag:nonlinear}
\end{align}
where $d_{m_f}^{(2)}$ denotes the dimensionless quadratic coupling to the fermion mass term, $d_g^{(2)}$ the quadratic coupling to the gluon field strength, and $\gamma_m$ the quark-mass anomalous dimension. $m_f$ denotes the fermion vacuum mass and $M_{\rm Pl}=1.22\times 10^{19}$~GeV the Plank mass. In deriving the fermion-mass bounds below, we will use the standard single-coupling-at-a-time benchmark and set $d_g^{(2)}=0$. 

In this work we focus on the quadratic coupling to electrons, $d_{m_e}^{(2)}$, and on the commonly considered symmetric and antisymmetric light-quark combinations, 
\begin{align}
    d_{\hat m}^{(2)} \equiv \frac{m_d d_{m_d}^{(2)}+m_u d_{m_u}^{(2)}}{m_d+m_u}\,,
    d_{\delta m}^{(2)} \equiv \frac{m_d d_{m_d}^{(2)}-m_u d_{m_u}^{(2)}}{m_d-m_u}\,,
    \label{eq:dquark}
\end{align}
where $d_{\hat m}^{(2)}$ enters through the average light-quark mass dependence of hadronic quantities while $d_{\delta m}^{(2)}$ controls the variation of the neutron--proton mass splitting.

The interactions in Eqs.~\eqref{lag:linear} and~\eqref{lag:nonlinear} radiatively induce  quartic self-interactions for $\phi$.  In addition to a possible tree-level interaction: $\lambda_0 \phi^4/4$, the one-loop effective potential is given by the Coleman–Weinberg form~\cite{Coleman:1973jx},
\begin{align}
    V_{\rm CW}(\phi) = -\frac{N_c}{64\pi^2}
    M_f^4(\phi) \left[ \ln\!\left(\frac{M_f^2(\phi)}{\mu^2}\right) -\frac{3}{2} \right],
    \label{eq:VCW}
\end{align}
with $N_c=1$ for leptons and $N_c=3$ for quarks. The $\phi$-dependent mass $M_f(\phi)$ gives $M_f(\phi)= m_f + y_f\phi$ for the linear case and $M_f(\phi)= m_f( 1+2\pi d_{m_f}^{(2)} \phi^2/M_{\rm Pl}^2)$ for the nonlinear case. Expanding Eq.~\eqref{eq:VCW} to fourth order in $\phi$ and neglecting the tiny mass correction from $\phi$ inside the logarithm would yield the induced quartic couplings 
\begin{align}
    \lambda_{\rm lin} &= -\sum_f
    \frac{N_c y_f^4}{16\pi^2} \left[ \ln\!\left(\frac{m_f^2}{\mu^2}\right) -\frac{3}{2} \right],
    \label{eq:linlambda}
    \\
    \lambda_{\rm quad} &= -\sum_f \frac{3N_c m_f^4}{2M_{\rm Pl}^4} \left(d_{m_f}^{(2)}\right)^2 \left[ \ln\!\left(\frac{m_f^2}{\mu^2}\right) -\frac{3}{2} \right],
    \label{eq:quadlambda}
\end{align}
for linear and quadratic interactions, respectively. Here, $\mu$ is the renormalization-group (RG) scale in the $\overline{\text{MS}}$ scheme.  We evaluate the fermion-induced quartic coupling as a threshold correction at $\mu = m_f$, corresponding to the scale at which the fermion is integrated out and the logarithm in the Coleman--Weinberg contribution is minimized. Below this threshold, the quartic coupling runs mainly through scalar self-interactions, $d\lambda/d\ln\mu = 9\lambda^2/(8\pi^2)$, which is numerically negligible given that the self-interaction coupling should be extremely small.

We should mention that for light-quark couplings, Eq.~\eqref{eq:quadlambda} should be regarded as a quark-level matching estimate. Since $u, d$ quarks are confined below $\Lambda_{\rm QCD} = 200 {\rm \, MeV}$, a fully controlled treatment requires matching the $\phi$-dependent quark masses onto the QCD vacuum energy, or equivalently onto chiral effective field theory. The resulting bounds are expected to remain parametrically indicative, while the precise coefficients require careful nonperturbative QCD computations, that will be done in a companion paper.

The effective coupling  receives contributions from both the renormalized tree-level term and the radiative corrections, $\lambda_{\rm eff}=\lambda_0+\lambda_{\rm in} (\lambda_{\rm quad})$, which can become $\mu$ independent under the one-loop RG-improved potential~\cite{Kastening:1991gv,Bando:1992np,Ford:1992mv}. In the absence of a fine-tuned cancellation against the renormalized $\lambda_0$, the contribution from $\lambda_{\rm lin}$ or $\lambda_{\rm quad}$  must not exceed the observational bounds on ULDM self-interactions.  Therefore, one can apply the stringent astrophysical and cosmological bounds of $\lambda_{\rm eff}$ to obtain  constraints on $\lambda_{\rm lin}$ and $\lambda_{\rm quad}$.  This provides a mapping from self-interaction bounds to constraints on the underlying ULDM couplings to fermionic fields.

\textbf{\textit{Constraints on ULDM self-interactions}}.
ULDM self-interactions are tightly constrained by a variety of astrophysical and cosmological observations, which we use as input for our analysis.  On cosmological scales, self-interactions modify the evolution of linear density perturbations through an effective sound speed, shifting the Jeans scale and suppressing structure formation below a characteristic cutoff. Requiring consistency with CMB anisotropies and large-scale structure (LSS) observations leads to a stringent upper bound on repulsive self-interactions~\cite{Cembranos:2018ulm},
\begin{align}\label{bound:CMB+LSS}
 \text{CMB+LSS}:\lambda_{\rm eff}<1.38\times 10^{-92}\left(\frac{m_\phi}{10^{-22}~\text{eV}}\right)^4,
\end{align}
valid for $m_\phi \gtrsim 10^{-24}~\mathrm{eV}$. Future CMB lensing measurements can further probe this effect by accessing smaller scales. As an indicative estimate, one may rescale Eq.~\eqref{bound:CMB+LSS} according to the expected sensitivity to the ULDM cutoff scale. A CMB-HD-like survey
could reach~\cite{Nguyen:2017zqu,Sehgal:2019nmk, Sehgal:2019ewc} 
\begin{align}
   \text{CMB-HD}: \lambda_{\rm eff}<1.26\times 10^{-94}\left(\frac{m_\phi}{10^{-22}~\text{eV}}\right)^4,
    \label{bound:CMB-HD+LSS}
\end{align}
which will be adopted  as an estimate, though a full dedicated likelihood analysis deserves considerations.

Self-interactions also modify the effective equation of state of ULDM. In contrast to cold DM with $w=0$, a quartic interaction induces a nonzero pressure scaling as $w \propto \lambda_{\rm eff}/m_\phi^4$. One may also derive the constraint by requiring that the background evolution remains close to that of cold DM with the bound $|w| \lesssim 10^{-3}$ at matter--radiation equality~\cite{Cembranos:2018ulm}. However, the bound is much weaker than from CMB anisotropies and LSS.

On astrophysical scales, quartic self-interactions can modify the structure of DM halos through the properties of solitonic cores governed by the Schr\"odinger--Poisson equations. Repulsive interactions (positive self-coupling) increase the core size, while attractive interactions (negative self-coupling) lead to more compact configurations.  For repulsive interactions, we found that the leading bounds arise from CMB and LSS observations. For attractive interactions, on the other hand,  observations of the DM distribution in galaxies, in particular in M87, impose strong constraints on a negative $\lambda_{\rm eff}$~\cite{Chakrabarti:2022owq} with an upper bound $-\lambda_{\rm eff}<2.34\times 10^{-94}(m_\phi/10^{-22}~\text{eV})^2$.  

In this Letter, we will  present  the fermion case for illustration purpose. We should  apply the general mapping for repulsive self-interactions since the fermion loop-induced $\lambda_{\rm lin}$ and $\lambda_{\rm quad}$ are positive. It turns out that such mapping can provide a powerful  probe of the underlying couplings between ULDM and fermionic fields.

\begin{figure}[t]
	\centering
\includegraphics[scale=0.54]{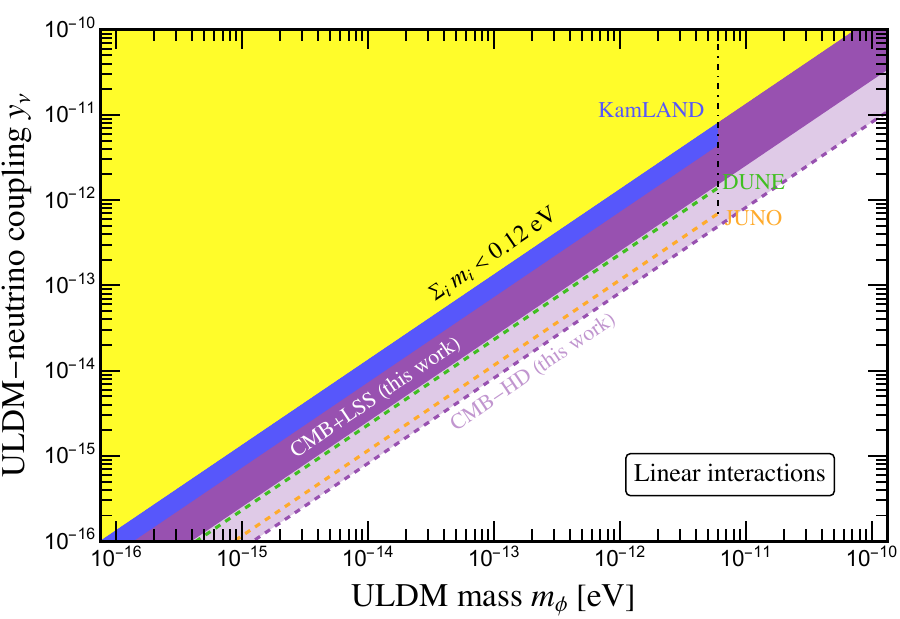}
	\caption{\label{fig:neutrino}
   Constraints on the linear ULDM coupling to neutrinos, parameterized by $y_{\nu}$ through Eq.~\eqref{lag:linear}, as a function of the ULDM mass $m_\phi$. The shaded dark and light regions show the bounds derived in this work from loop-induced ULDM self-interactions: the dark region corresponds to current CMB+LSS limits, Eq.~\eqref{bound:CMB+LSS}, while the light region denotes the forecast sensitivity of CMB-HD, Eq.~\eqref{bound:CMB-HD+LSS}. For comparison, the leading existing limits come from the cosmological bound on the sum of neutrino masses, $\sum_i m_i < 0.12~\mathrm{eV}$, taken from Planck~\cite{Planck:2018vyg}, and from long-baseline reactor-neutrino oscillation experiments~\cite{Krnjaic:2017zlz,Dev:2020kgz}.
    }
\end{figure}

\textbf{\textit{Bounds on ULDM--fermion couplings}}.
The loop-induced self-interaction derived above provides a direct and model-independent mapping between the ULDM-fermion couplings and the effective quartic interaction. While we found that these bounds do not generically supersede the strongest existing limits on linear ULDM couplings to charged fermions, they can provide a complementary, model-independent constraint. For instance,  consider the linear coupling to electrons, where $\lambda_{\rm lin} = 3y_e^4/(32\pi^2)$. It must satisfy the  bound given in Eq.~\eqref{bound:CMB+LSS}. For $m_\phi=10^{-22}\,\mathrm{eV}$, the recast bound on the electron linear coupling is $y_e\lesssim 3\times 10^{-23}$. This estimate shows that the bound on $y_e$ is only an order of magnitude weaker than the MICROSCOPE limit~\cite{Berge:2017ovy,Hees:2018fpg}, and is comparable to that obtained from Eöt-Wash experiment~\cite{Su:1994gu,Smith:1999cr,Schlamminger:2007ht,Wagner:2012ui}.

The impact of the self-interaction bounds is particularly pronounced in the neutrino sector, where a broader range of parameter space is still viable. ULDM--neutrino interactions can induce observable distortions in neutrino oscillation probabilities in solar~\cite{Berlin:2016woy} and reactor~\cite{Krnjaic:2017zlz,Dev:2020kgz} experiments. In particular, the induced contribution to neutrino masses, $\delta m_\nu = y_\nu \phi$, affects both the absolute masses and the mixing parameters. A shift in the masses arises from flavor-diagonal couplings, while time-dependent modulations of the mixing angles require flavor off-diagonal entries in $y_\nu$. The ULDM background can also induce time variations in the CP-violating phase~\cite{Losada:2021bxx}, although current measurements are not yet sensitive to this effect.

We show in Fig.~\ref{fig:neutrino} the resulting bounds on the linear ULDM--neutrino coupling as a function of the ULDM mass $m_\phi$. We only show the mass range between $10^{-16}$~eV and $10^{-10}$~eV for better visualization of the different bounds, but applying  our bounds to lower and higher masses is straightforward. The shaded region with $\sum_i m_i<0.12$~eV is obtained by requiring that the sum of neutrino masses should not exceed the current bounds from cosmic observations~\cite{Planck:2018vyg}. Probes of ULDM for masses up to $m_\phi\simeq 10^{-12}-10^{-11}$~eV can be achieved by long-baseline reactor neutrino experiments such as KamLAND~\cite{KamLAND:2013rgu}, JUNO~\cite{JUNO:2015zny}, and DUNE~\cite{DUNE:2015lol}.
The  shaded region referred to as KamLAND corresponds the exclusion limits  based on the KamLAND experiment~\cite{Krnjaic:2017zlz}, while the  dashed lines referred to as DUNE and JUNO represent the forecast sensitivity from the DUNE~\cite{Dev:2020kgz} and JUNO experiments~\cite{Losada:2021bxx}, respectively. 

\begin{figure*}[t!]
    \centering
    \begin{minipage}[t]{0.33\textwidth}
        \centering
        \includegraphics[width=\linewidth]{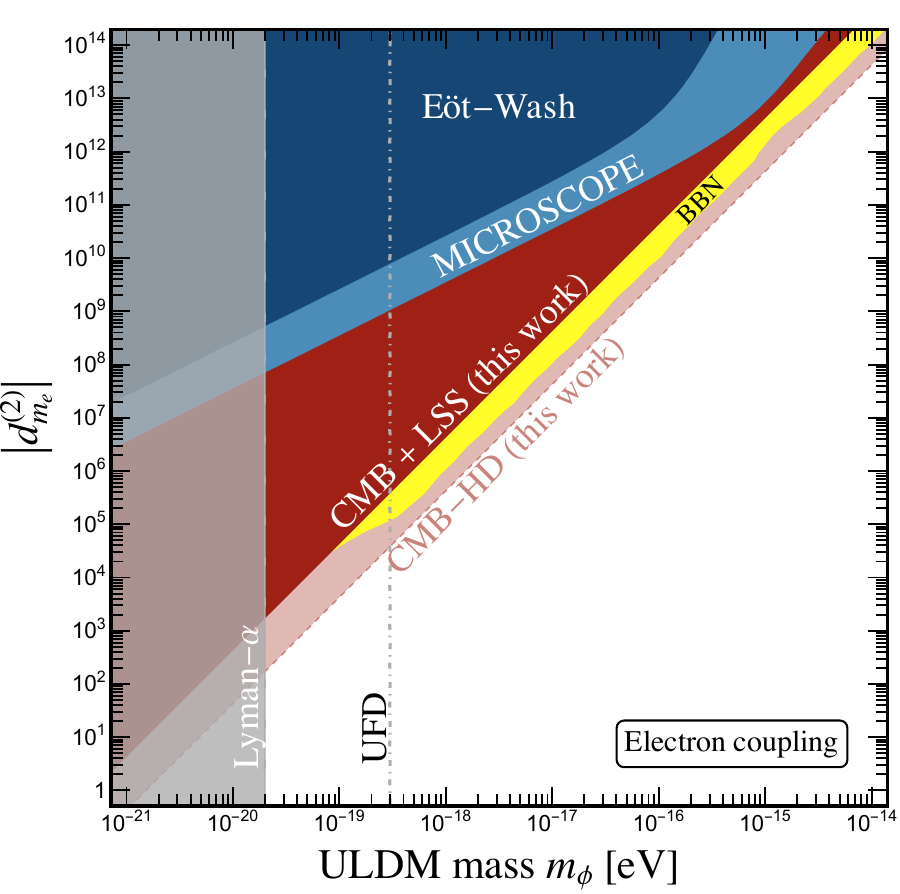}
    \end{minipage}
    \hfill
    \begin{minipage}[t]{0.33\textwidth}
        \centering
        \includegraphics[width=\linewidth]{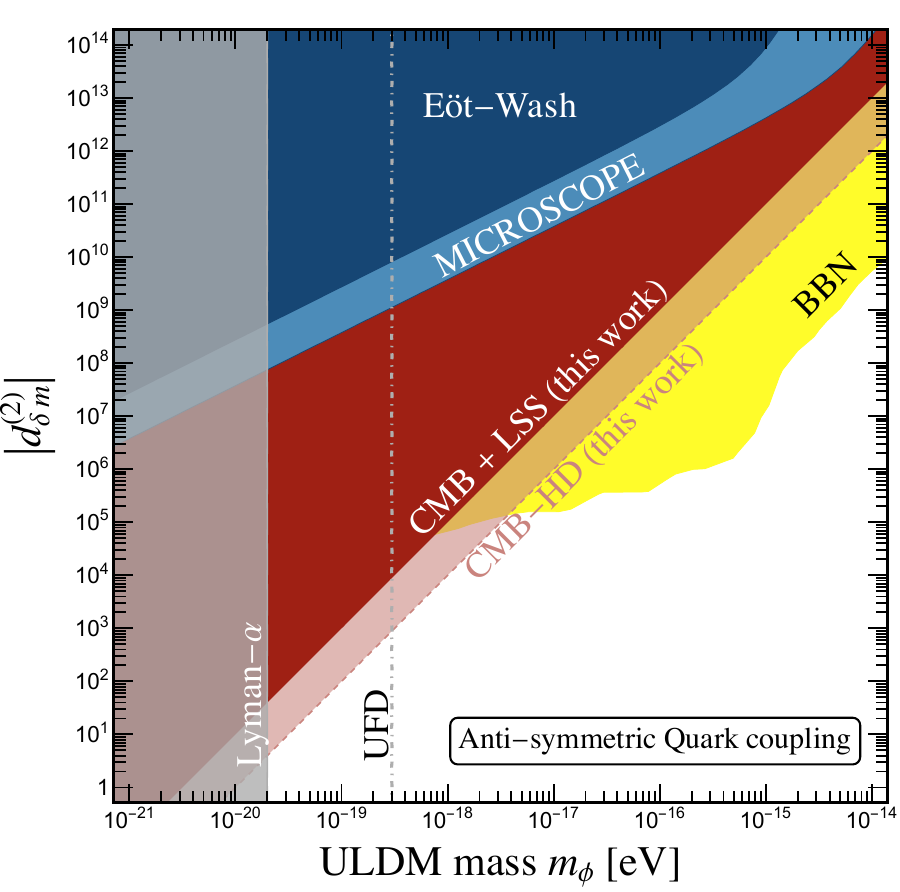}
    \end{minipage}
    \hfill
    \begin{minipage}[t]{0.33\textwidth}
        \centering
        \includegraphics[width=\linewidth]{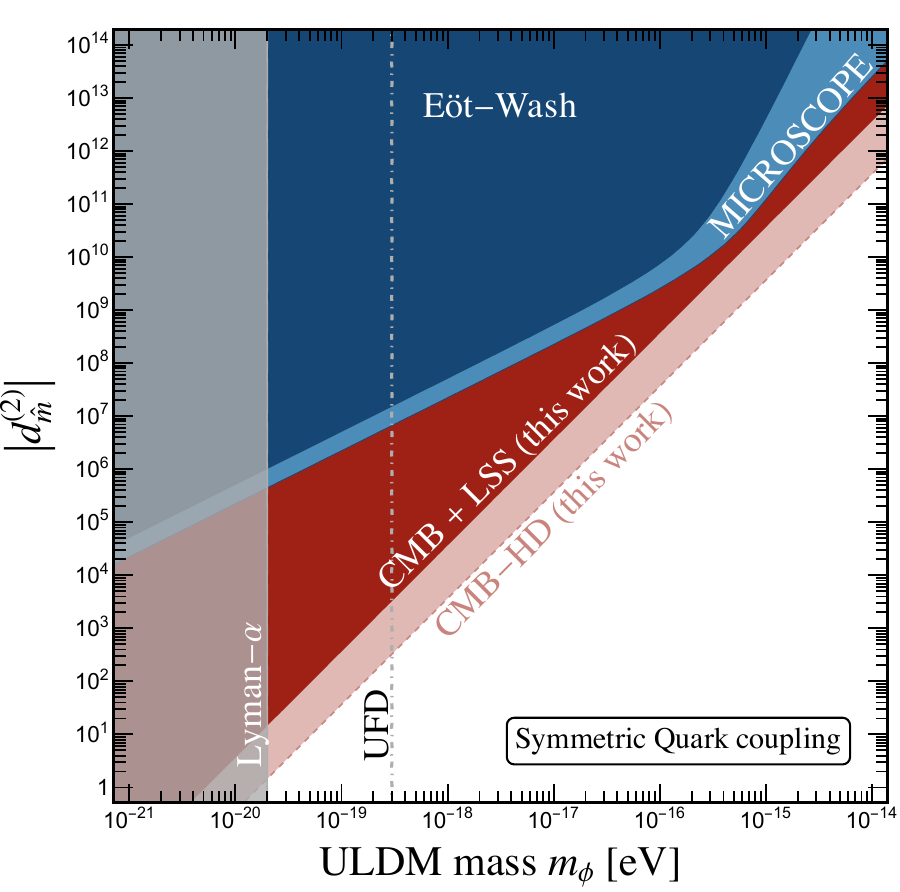}
    \end{minipage}
	\caption{ Constraints on quadratic ULDM couplings as a function of the ULDM mass $m_\phi$, shown for (left) electrons $d^{(2)}_{m_e}$, (center) quark anti-symmetric coupling $d^{(2)}_{\delta m}$, and (right) quark symmetric coupling $d^{(2)}_{\hat m}$; see Eqs.~\eqref{lag:nonlinear} and \eqref{eq:dquark}. The red shaded regions show the bounds derived in this work from loop-induced ULDM self-interactions: dark red corresponds to the current reach of CMB+LSS limits~\cite{Cembranos:2018ulm}, while light red denotes the projected sensitivity of future CMB-HD-like surveys~\cite{Sehgal:2019ewc}, with dashed contours indicating the corresponding projections. These bounds are model-independent and universal within the EFT considered here. For comparison, dark and light blue regions denote equivalence-principle bounds from E\"ot-Wash~\cite{Schlamminger:2007ht,Wagner:2012ui} and MICROSCOPE~\cite{Touboul:2017grn}, respectively. Yellow regions indicate BBN constraints~\cite{Bouley:2022eer}, which depend on the early-universe evolution of the ULDM field. Gray regions show coupling-independent bounds from Lyman-$\alpha$ forest measurements~\cite{Rogers:2020ltq} and ultrafaint dwarf galaxies (UFD)~\cite{Dalal:2022rmp}.
}
    \label{fig:quadratic}
\end{figure*}

The bounds derived from neutrino oscillation experiments are generally weaker than from the self-interaction. Consequently, repulsive ULDM self-interactions can limit the probe of ULDM through neutrino oscillation experiments. We should emphasize that the bounds derived from the oscillation measurements of mixing angle modulation are valid only for flavor-violating couplings, i.e., with off-diagonal entries in $y_\nu$, while the self-interaction bound acts universally upon all $y_\nu$ entries. This feature suggests that the constraint proposed in this Letter also applies to any linear ULDM-matter interactions that exhibit flavor-changing processes such as in  charged-lepton and quark sectors, which remains unexplored thus far.

We now turn to quadratic couplings between ULDM and fermions. Such interactions are well motivated in scenarios where a symmetry suppresses the linear term, and have recently attracted considerable attention due to their distinct phenomenology and their ability to evade leading fifth-force constraints. Despite the absence of a leading long-range force, quadratic couplings inevitably induce a quartic self-interaction for the ULDM field through fermion loops. As shown in Eq.~\eqref{eq:quadlambda}, the resulting effective coupling scales as $\lambda_{\rm quad}\propto (d_{m_f}^{(2)})^2$, so that observational bounds on the quartic self-interaction can be parametrically more powerful and sensitive than in the linear case  where the loop-induced quartic coupling arises at fourth order.

In Fig.~\ref{fig:quadratic}, we show the bounds on the quadratic coupling to electrons, $d_{m_e}^{(2)}$, as well as $d^{(2)}_{\delta m}$ and $d^{(2)}_{\hat m}$ defined through Eq.~\eqref{eq:dquark}. The dark and light blue shaded regions show existing equivalence-principle bounds from E\"ot-Wash~\cite{Schlamminger:2007ht,Wagner:2012ui} and MICROSCOPE~\cite{Touboul:2017grn}, respectively.
The yellow shaded region denotes the constraint from BBN, during which the coherent ULDM background induces shifts in fundamental constants and hence affects weak interaction rates~\cite{Bouley:2022eer}. 
This leads to significant constraints on $d_{m_e}^{(2)}$ and $d^{(2)}_{\delta m}$, while  no meaningful bound is obtained for  $d^{(2)}_{\hat m}$  whose impact enters only indirectly through hadronic quantities such as the neutron axial coupling. While the BBN bounds are typically stronger than equivalence-principle tests, it is important to emphasize that the BBN constraint depends on the cosmological evolution of the ULDM field. In particular, it assumes that ULDM is already present and coherently oscillating during the BBN epoch with an amplitude fixed by the present-day DM abundance. This assumption can be relaxed in scenarios with late-time production, rendering the resulting bounds intrinsically model dependent.
The gray vertical bands indicate constraints that are essentially independent of the coupling strength. Lyman-$\alpha$~\cite{Irsic:2017yje,Armengaud:2017nkf,Kobayashi:2017jcf,Rogers:2020ltq} observations exclude sufficiently small ULDM masses due to the suppression of small-scale structure, while ultrafaint dwarf galaxies (UFD)~\cite{Dalal:2022rmp} provide complementary limits in a similar mass range. 

The red shaded regions show the bounds derived in this work from loop-induced self-interactions. The dark (light) red shading corresponds to current CMB+LSS limits (future CMB-HD+LSS sensitivity), obtained by translating the self-interaction bounds from Eqs.~\eqref{bound:CMB+LSS} and~\eqref{bound:CMB-HD+LSS} into constraints on $d_{m_f}^{(2)}$. Across most of the parameter space, these bounds are comparable to or stronger than BBN where applicable, and otherwise provide the leading constraints, exceeding equivalence-principle limits by several orders of magnitude. By contrast with BBN, our constraints rely solely on present-day astrophysical and cosmological limits on ULDM self-interactions and on the irreducible loop-induced quartic coupling. They are therefore insensitive to the ULDM production mechanism and provide a robust, model-independent probe of quadratic ULDM couplings.

The implications of our results for phenomenological studies and detection proposals targeting quadratically coupled ULDM are significant. Recent studies of symmetry-protected scenarios such as the quadratic twin~\cite{Delaunay:2025pho}, have emphasized that sizable quadratic couplings to electrons can remain phenomenologically viable. However, Fig.~\ref{fig:quadratic} indicates that this conclusion must be substantially revised once loop-induced self-interactions are taken into account. Even if the naturalness problem is alleviated by symmetry arguments, such as the twin-Higgs mechanism~\cite{Chacko:2005pe}, the loop-induced self-interaction is not generically removed by the same symmetry. As a result, once self-interactions are taken into account, a large fraction of the previously viable parameter space will be excluded.

Finally, we emphasize that the argument presented in this Letter is general. Any interaction that contributes to the scalar effective potential will induce self-interactions at loop level, and the resulting constraints can therefore be extended to other ULDM candidates and to couplings involving bosonic degrees of freedom. This establishes self-interaction bounds as a robust and broadly applicable probe of ULDM interactions with the SM or even hidden particles.

\textbf{\textit{Conclusion}}. We have shown that bounds on ULDM self-interactions can be directly translated into constraints on ULDM couplings to SM fields through the loop-induced quartic interactions generated by these couplings. Using this connection, we  found that, for the first time, both linear and nonlinear interactions are subject to stringent constraints. In particular, the resulting bounds provide the leading limits on linear ULDM--neutrino couplings and on quadratically coupled ULDM across wide regions of parameter space. While we have presented explicit results for quadratic couplings to electrons and light quarks, parametrically similar bounds are expected for other fermions and for bosonic couplings, since the underlying mechanism is generic. Our analysis has focused on flavor-diagonal interactions; however, the same framework can also be extended to flavor non-diagonal couplings,  which  is  expected to provide strong bounds on  quark- or lepton-flavor changing processes and distorted neutrino oscillation probabilities. Our results establish ULDM self-interaction bounds as a powerful test for ULDM--matter interactions. More generally, they demonstrate that quantum-induced effects, when combined with astrophysical and cosmological observations, can impose robust and far-reaching constraints on the microscopic interactions of ULDM.

\textbf{\textit{Acknowledgements}}. We would like to thank Helmut Eberl, Josef Pradler, and Ryosuke Sato for discussions.	S.-P.~Li would also like to thank Shinya Kanemura for financial support for transition from the University of Osaka to MBI, OAW, when this work was initiated. M.~Aghaie is supported by JSPS KAKENHI Grant Number 24H02244. The authors contribute equally to this work.

\bibliographystyle{JHEP}
\bibliography{Refs}

\end{document}